\documentclass[12pt]{article}

\usepackage{graphicx}
\usepackage{amssymb}
\usepackage{amsmath}
\usepackage{color}
\usepackage{url}
\usepackage{listings}
\usepackage{hyperref}

\newcommand{\definition}[1]{\textit{#1}}

\title{Programming Languages for Scientific Computing}
\author{Matthew G. Knepley\\
\small Computation Institute\\[-0.8ex]
\small University of Chicago, Chicago, IL\\
\small \texttt{knepley@ci.uchicago.edu}\\
}

\begin{document}
\maketitle

\section{Introduction}

Scientific computation is a discipline that combines numerical analysis, physical understanding, algorithm development,
and structured programming. Several yottacycles per year on the world's largest computers are spent
simulating problems as diverse as weather prediction, the properties of material composites, the behavior of
biomolecules in solution, and the quantum nature of chemical compounds~\cite{ScalesReport}. This article is intended to
review specfic languages features and their use in computational science. We will review the strengths and weaknesses of
different programming styles, with examples taken from widely used scientific codes.

This article will not cover the broader range of programming languages, including functional and logic languages, as
these have, so far, not made inroads into the scientific computing community. We do not cover systems with sophisticated
runtime requirements, such as Cilk~\cite{Cilk95}, since this is currently incompatible with high performance on cutting edge
hardware. For this reason, we also ignore transactional memory, both software and hardware. We also will not discuss the
particular capabilities of software libraries in detail. Particular libraries will be used as examples, in order to
highlight advantages and disadvantages of the programming paradigm, but no exhaustive presentations of their
capabilities, strengths, or weakenesses will be given.

\section{Brief Overview of Language Characteristics}

We begin our discussion with \definition{imperative} languages, like C and Fortran, meaning languages where the programmer
explicitly tells the computer what to do at each step. The computation is built from variables, which hold values, and
functions which compute the value of a variable based upon the input values of other variables. For instance,
important functions for scientific computing are arithmetic functions, like division, and linear algebraic functions,
like matrix multiplication. The principal advantage of these imperative languages over simpler systems, such as Excel, is the
ability to flexibly combine these basic elements.

In C and Fortran 90, groups of related variables can be combined together in a \definition{structure}, which allows them
to be passed as a unit to functions. This both improves code readability and decreases its conceptual
complexity. For example, a customer structure could store a customer's name, account number, and outstanding balance.
\lstset{language=C,basicstyle=\small,stringstyle=\ttfamily,morekeywords={MPI_Comm,KSP,PC}}
\begin{lstlisting}
struct customer {
  char *name;
  int   acct;
  float balance;
};
\end{lstlisting}
Similarly, functions may call other functions, or themselves recursively, in order to simplify the description of the
operation. For example, the merge sort algorithm works by first sorting each half of an array, and then merging together
these smaller sorted arrays into a completely sorted array.
\begin{lstlisting}
void mergeSort(int array[], int arrayLength) {
  int halfLength = arrayLength/2;

  if (arrayLength < 2) return;
  mergeSort(&array[0], halfLength);
  mergeSort(&array[halfLength], halfLength);
  merge(&array[0], &array[halfLength]);
}
\end{lstlisting}
Using these mechanisms, just amounting to the introduction of hierarchical organization to simple code elements, the
complexity of large codes can be drastically reduced.

\definition{Object-Oriented} languages, such as C++ and Python, allow a further level of combination. Data can be
grouped together with the functions which operate on it, into a super-structure called an \definition{object}. This can
be useful for organizing the action on groups of data. For example, we can augment our customer example with methods
which change the account number or debit the account, where now we declare a \definition{class} which describes a type of object.
\lstset{language=c++}
\begin{lstlisting}
class customer {
  char *name;
  int   acct;
  float balance;
public:
  void debit(float amount) {
    this->balance -= amount;
  };
  void changeAccount(int acct) {
    this->acct = acct;
  };
}
\end{lstlisting}
However, this organization can also be accomplished in standard C by passing the structure as an argument.
\lstset{language=c}
\begin{lstlisting}
void debit(struct customer *this, float amount) {
  this->balance += amount;
}
\end{lstlisting}

Another organizational strategy is to give \definition{types} to variables. In C and Fortran, this tells the compiler
how much space to use for a variable, such as 4 bytes for a \lstinline!long int! in C. Structures are also types, built
out of smaller types, as are classes. In some languages, such as C, C++, and Fortran, the type of every variable must be
specified before the program is run, which is called \definition{static typing}. In contrast, Python, Ruby, and Perl
allow the type of a variable to change at runtime depending on what kind of value is stored in it, which is called
\definition{dynamic typing}. Dynamic typing makes code smaller and easier to write, but the code is harder for the
compiler to optimize and can sometimes be harder to understand without types to guide the reader.

Object-oriented languages very often have collections of similar functions that operate differently depending on the
type of argument provided, or the type of object associated with the function since the object is understood as a silent
first argument. For example,
\lstset{language=c++}
\begin{lstlisting}
class circle {
  float radius;
public:
  float area() {
    return PI*this->radius*this->radius;
  };
}

class triangle {
  float base, height;
public:
  float area() {
    return 0.5*this->base*this->height;
  };
}
\end{lstlisting}
the \lstinline!area()! function will behave differently when called with a circle object, rather than a
triangle. Choosing a specific function, or \definition{method dispatch}, based upon the types of its arguments is called
\definition{polymorphism}. A programmer might want two classes to share many functions and data, but differ in a few
respects. The \definition{inheritance} mechanism allows one class to behave exactly as another, unless that behvior is
explicitly redefined.

In languages with static typing, it can be useful to write functions which have the same form for a range of types, just as
they would look in a dynamically typed language. This mechanism is called \definition{genericity}, and the specific
strategy used in C++ is \definition{templating}. Templates allow a placeholder, often \lstinline!T!, to be replaced by
the specific type of an argument when the code is compiled. Thus many versions of the function are generated, a process
called \definition{template instantiation}, one for each different type of argument.

\section{Single language Codes}
  \subsection{Imperative Programming}\label{sec:imperative}

\paragraph{Advantages} The still dominant paradigm for both application code and libraries in scientific computing is a
single language code base in a well-established imperative language such as C or FORTRAN 77 (F77). These languages have
several notable advantages over more sophisticated alternatives when applied to numerical algorithms. First and
foremost, they can be made performant by a mildly proficient user, and the ease of achieving good performance comes from
several language features. Both C and Fortran are very similar to the underlying assembly code into which they are
compiled. Thus, it is not only obvious to users how a given routine will be executed, but also obvious to the
compiler. This correspondence makes it much easier to create routines that compilers can optimize well. The simple
execution model for C and F77 also makes inspection of the code by an outside user possible. More complex constructs,
such as templates and deep inheritance hierarchies, can obscure the actual execution even while making the intent
clearer. Moreover, the state of the computation and data structures can be easily seen in a debugger, whereas more
complex constructs and execution environments often hide this information.

Simplicity in execution also translates to greater ease in using debugging and profiling tools. Major debugging tools
such as gdb, idb, totalview, and \href{http://valgrind.org/}{valgrind}~\cite{valgrind-web-page} have excellent support for C and F77. They do
support higher level features, but there can be inconsistencies, especially with template instantiation, that cause some
information to be unavailable. This caveat also applies to profiling tools. Simplicity in binary interface definition
means that C and F77 are especially easy to interface with other languages and environments. Symbols are not
\definition{mangled}, or made unique using complex names, so matching ones can be easily created in another system.
Function parameter passing is also unambiguous. This makes C the language of choice when defining a
\definition{foreign function} interface for a higher level language, that is an interface which allows functions in one
language to be called from another such as C.

\paragraph{Disadvantages} A price is paid, however, for the simplicity of these languages. The size of source code for
equivalent tasks is quite often more than an order of magnitude larger than for object oriented or functional languages. The
user must write code for method dispatch instead of using polymorphism, write separate routines for many types instead
of using templates, produce basic data structures which are not part of core libraries, and in general reproduce many
of the mechanisms built into higher level languages, as described below. In particular, the rich standard libraries of
many OO and functional languages save considerable special purpose code.

\lstset{language=c++}
One of the most severe omissions in C and F77 is that of flexible namespaces for identifiers, types, and functions. The
absence of hierarchical namespaces for symbols, such as \lstinline!namespace! in C++ or \definition{dot} notation in
Python, results in comically long identifier names, and rampant problems with clashing symbol names when linking together
different scientific libraries. A second problem is the need for manual memory management of all structures, or for F77
static declaration of memory up front. In C++, when objects are declared in an inner scope such as a function body or
for loop, they are automatically created upon entry and destroyed on exit from that scope. These are called
\definition{automatic objects}, and arrays can also be defined this way. In C, the creation and destruction must be
managed by hand, which may be complicated when for instance error conditions arise. Lastly, there
are no language facilities for \definition{introspection}, determination of code structure at runtime, as there are in
C++ or Python. At best, we can use the dynamic loading infrastructure to search for library symbols, but cannot
determine which types, functions, or structures are defined in a library without making separate, configuration tests
outside the language itself. This usually results in fantastic complication of the build process.

\lstset{language=fortran}
\paragraph{Example} Perhaps the most successful software library written in this paradigm are the BLAS
library~\cite{BLAS79}, dating from 1979, and \href{http://www.netlib.org/lapack/}{LAPACK}~\cite{LAPACK90} libraries for linear algebra, first released in
February 1992, for linear algebra. They are both numerically robust and extremely efficient, and used in almost every
serious numerical package. The internals are so easily understood, being written in simple F77, that they are often
copied wholesale into application code without the use of the library itself. However, they suffer from a classic
problem with scientific software of this type, lack of \definition{data encapsulation}. The data structures upon which
the operations, such as matrix-matrix multiplication, operate are specified directly in the library API. Thus the layout
of dense matrices is given in the interface and cannot be changed by the implementation. For example, the calling
sequence for double precision matrix-matrix multiplication in BLAS, a workhorse of scientific computing, is
\begin{lstlisting}
SUBROUTINE DGEMM(TRANSA, TRANSB, M, N, K, ALPHA, A, LDA,
                 B, LDB, BETA, C, LDC)
*     .. Scalar Arguments ..
      DOUBLE PRECISION ALPHA,BETA
      INTEGER K,LDA,LDB,LDC,M,N
      CHARACTER TRANSA,TRANSB
*     ..
*     .. Array Arguments ..
      DOUBLE PRECISION A(LDA,*),B(LDB,*),C(LDC,*)
\end{lstlisting}
Here the multiply is prescribed to operate on a dense array of doubles \lstinline!A! with a row stride of
\lstinline!LDA!. This limitation complicated the implementation of an efficient distributed memory version of the
library, and led to the introduction of \href{http://code.google.com/p/elemental/}{Elemental}~\cite{Elemental2012} which uses a more
favorable data distribution, especially for smaller sizes. It has also prevented the fusion of successive operations,
which could result in data reuse or latency hiding, greatly improving the efficiency of the library.

  \subsection{Object Orientation}\label{sec:oo}

\paragraph{Advantages} Object Orientation (OO) is a powerful strategy for data encapsulation. Objects are structures
that combine data and functions which operate on that data. Although this can clearly be accomplished in C using
\lstinline!struct!s and function pointers, many languages have builtin support for this, including Objective C, C++,
C\#, and Python. This kind of encapsulation encourages the programmer to produce \definition{data structure neutral}
interfaces~\cite{dsneutral}, as opposed to those in LAPACK. Combined with polymorphism, or function dispatch based upon
the argument types, we can write a single numerical code that uses different algorithms and data structures based upon
its input types~\cite{smith98}. This, in a nutshell, is the current strategy for dealing with the panoply of modern
architectures and problem characteristics for scientific simulation. It should also be noted that all the OO languages
mentioned above provide excellent namespacing facilities, overcoming another obstacle noted in Section~\ref{sec:imperative}.

The essential features of OO organization, encapsulation and dynamic dispatch, can be emulated in C at the cost of many
more lines of code. Early C++ compilers did just this by emitting C rather than object code. Moreover, languages such as
C++ and Java have removed some of the dynamism present in Objective C and C OO frameworks. We will show an example of
this below.

\paragraph{Disadvantages} The downsides of object oriented organization have to do with controlling code complexity, the
original motivation for the introduction of OO structures. The true measure of code complexity is ease of understanding
for an outside observer. There can be a temptation to create deep object hierarchies, but this tends to work against
both code readability and runtime flexibility as illustrated below. For numerical code especially, it is common to
introduce operator overloading. This can improve readability, however transparency of the performance cost is lost,
which often results in very slow application code, unacceptable in most simulation environments.

\paragraph{Examples} \href{http://www.mcs.anl.gov/petsc/}{PETSc}~\cite{petsc-user-ref,petsc-web-page} and \href{http://trilinos.sandia.gov/}{Trilinos}~\cite{trilinos:overview,trilinos:homepage}
are two popular packages which can solve sparse systems of nonlinear algebraic equations in parallel. A common case for
which these libraries use OO techniques to control complexity is the choice among a dizzying range of iterative solvers
and preconditioners.

In PETSc, a Krylov Subspace solver (\lstinline!KSP!) object acts as an abstract base class in C++. However, the key
difference is that instantiation of the subtype is done at runtime,
\lstset{language=C,basicstyle=\small,stringstyle=\ttfamily,morekeywords={MPI_Comm,KSP,PC}}
\begin{lstlisting}
  MPI_Comm comm;
  KSP      ksp;
  PC       pc;

  KSPCreate(comm, &ksp);
  KSPGetPC(ksp, &pc);
  /* Generally done with command line options */
  KSPSetType(ksp, "gmres");
  PCSetType(ksp, "ilu");
\end{lstlisting}
and we see that the Trilinos equivalent in C++ is almost identical.
\lstset{language=C++,basicstyle=\small,stringstyle=\ttfamily,morekeywords={AztecOO}}
\begin{lstlisting}
  Teuchos::RCP<Epetra_RowMatrix> A;
  Epetra_Vector LHS, RHS;
  Epetra_LinearProblem Problem(&*A,&LHS,&RHS);
  Ifpack Factory;
  Teuchos::RCP<Ifpack_Preconditioner> Prec =
    Teuchos::rcp(Factory.Create("ILU", &*A, 1));
  AztecOO Solver(Problem);

  Solver.SetAztecOption(AZ_solver, AZ_gmres);
  Solver.SetPrecOperator(&*Prec);
\end{lstlisting}
Trilinos and PETSc make the same decision to trade language support for runtime flexibility. In packages like dealII and
FEniCS, each linear solver is instantiated as a separate type which all derive from an abstract base type. Naively, this
strategy would force the user to change the application code in order to try a different solver. The Factory
Pattern~\cite{GammaHelmJohnsonVlissides95} is often used to alleviate this difficulty. Both Trilinos and PETSc also use
factories to organize instantiation.

However, two related problems arise. First, if the solver object is defined by a single concrete type, changing a given
solver nested deeply within a hierarchical solve becomes prohibitively complex. Both solver objects above can change the
concrete type ``on the fly''. This ability is key in multiphysics simulations where already complex solvers are combined
and nested. Second, accessing the concrete solver type would now involve downcasts that may fail, littering the code
with obtrusive checks. In PETSc, each concrete type has an API which is ignored by other types. Thus,
\lstset{language=C,basicstyle=\small,stringstyle=\ttfamily,morekeywords={MPI_Comm,KSP}}
\begin{lstlisting}
  KSPGMRESSetRestart(ksp, 45);
  KSPChebychevSetEigenvalues(ksp, 0.9, 0.1);
  PCFactorSetLevels(pc, 1);
  PCASMSetOverlap(pc, 2);
\end{lstlisting}
will execute without error regardless of the solver type, but will set eigenvalue bounds if the user selected the
Chebychev method. Trilinos uses a bag of parameters,
\lstset{language=C++,basicstyle=\small,stringstyle=\ttfamily,morekeywords={AztecOO}}
\begin{lstlisting}
  Teuchos::ParameterList List;

  List.set("fact: drop tolerance", 1e-9);
  List.set("fact: level-of-fill", 1);
  List.set("schwarz: combine mode", "Add");
  Prec->SetParameters(List);
\end{lstlisting}
however this sacrifices type safety for the arguments, and can also result in aliasing of argument names.

  \subsection{Code Generation}\label{sec:generation}

\paragraph{Advantages} Performance has always been a primary concern for numerical codes. However, the advent of new,
massively parallel architectures, such as the \href{http://www.nvidia.com/object/fermi_architecture.html}{Nvidia Fermi}~\cite{NvidiaFermi}
or \href{http://www.intel.com/content/www/us/en/architecture-and-technology/many-integrated-core/intel-many-integrated-core-architecture.html}{Intel MIC}~\cite{IntelMIC}, while
providing much more energy efficient performance, has greatly increased the penalty for suboptimal code. These chips
have vector units accomodating from 4 to 16 double precision operations, meaning that code without vectorization will
achieve no more than 25\% of peak performance, and usually much less. They also increase the imbalance between flop rate
and memory bandwidth or latency, so that thousands of flops can be needed to cover outstanding memory references. GPUs
in particular have very high memory latency coupled with a wide bus, making the memory access pattern critical for good
performance. In addition, the size of fast cache memory per core has shrunk dramatically, so that it cannot easily be
used to hide irregular memory access.

The strategies for mitigating these problems are familiar, and include
tiling~\cite{AbuSufahKuckLawrie81,GuoBikshandiFraguelaGarzaranPadua08}, redundant computation, and reordering for
spatial and temporal memory locality~\cite{GKKS-PCFD99,Strout04}. The CUDA language incorporates two of the most
important optimizations directly into the language: vectorization and memory latency hiding through fast context
switch~\cite{NvidiaFermi}. In CUDA, one writes \definition{kernels} in a Single Instruction Multiple Thread (SIMT) style,
so that vector operations are simple and explicit, in contrast to the complicated and non-portable compiler intrinsics
for the Intel MIC. These kernel routines may be swapped onto a processor using an extremely fast context switch,
allowing memory latency in one kernel to be hidden by computation in others. However, in CUDA itself, it is not possible
to express dependencies among kernels. \href{http://www.khronos.org/opencl/}{OpenCL}~\cite{OpenCLStandard} has preserved these essential
features of CUDA, and also achieves excellent performance on modern hardware.

It is, however, unlikely that these kernels can be coded by hand for scientific libraries. Even should the model,
discretization, coefficient representation, and solver algorithm be fixed, the kernel would still have to take account
of the vector length on the target processor, memory bus width, and available process local memory. We are not
describing merely tuning a small number of parameters describing the architecture, as for instance in
\href{http://math-atlas.sourceforge.net/}{Atlas}~\cite{atlas-web-page}, but algorithm reorganization at a high level, as shown in the examples.

\paragraph{Disadvantages} The pricipal disadvantage of automatically generated code are weak support in the build
toolchain. In contrast to C++ templates, more exotic methods of code generation require outside tools, usually separate
files, and are not easily incorporated into existing build system, especially for large projects. A very hopeful
development, however, is the incorporation in OpenCL of compilation as a library call. Thus kernel generation,
compilation, and linking can take place entirely within a running application, much like the template version.

However code is generated, care must be taken that the final output can be read by the user, and perhaps improved. A
major disadvantage of templates is that it prevents the user from directly inspecting the generated code. Without
readable code, the user cannot inspect the high level transformations which have been used, correct simple errors for
new environments, insert specialized code for new problems, and in general understand the system. Code generators should
strive to provide easy access for the user to generated source, as shown in the FEniCS package, while seamlessly
integrating the result into existing build architectures.

\paragraph{Examples} The \href{http://developer.nvidia.com/thrust}{Thrust}~\cite{thrust} package from Nvidia uses the C++ template
mechanism to generate CUDA kernels for common functional operations such as \lstinline!map!, \lstinline!transform!, and
\lstinline!reduceByKey!. Most transformations here amount to intelligent blocking and tiling, and are well suited to
this mechanism. Even higher level generation is used by both \href{http://www.spiral.net/}{Spiral}~\cite{Puschel05} and
\href{http://www.fftw.org/}{FFTW}~\cite{FrigoJohnson93}. The algorithm is broken down into smaller components, for FFTW these are ``codelets''
and Spiral produces another low-level language. A particular instantiation of the algorithm can be composed of these
pieces in many different ways. Partial implementations are constructed, run, and timed. This real time evaluation guides
the construction of the final implementation for the given problem.

  \subsection{Generiticity and Templating}\label{sec:generic}

\paragraph{Advantages} By far the most popular type of code generation technique employed in scientific computing is C++
templates. It gives users the ability to hardwire constants into a piece of code, allowing the compiler to fold them and
perform loop unrolling optimizations, without sacrificing flexibility in the code base or using convoluted preprocessing
schemes. It is also possible to write generic operations, independent of the data type on which they operate, but still
have them properly type check. This can make the code base much smaller, as separate routines for different types are
unified, and is the inspiration behind the \href{http://www.sgi.com/tech/stl/}{Standard Template Library}~\cite{Stepanov09,STLGuide}. Moreover, all
this can be done without changing the normal toolchain for C++ use, including compilation, profiling and debugging.

%% Paragraph on expression templates. MTL. Eigen.

\paragraph{Disadvantages} While templates are integrated into the normal C++ workflow, unfortunately the product of
template expansion is not available to the user. Thus, they cannot inspect the particular optimizations which are
performed or specialize it by adding code for a specific instance (although they can use the \definition{template
specialization} mechanism). Compile time also greatly increases with templates, becoming problematic for large code
bases. In addition, the type safety of templates is enforced at the instantiation point which can be very far removed
from the use location in the code. This very often results in inpenetrable, voluminous error messages that stretch for
hundreds of thousands of lines. The failure of \textit{concepts} to enter the C++ standard~\cite{Siek2012} seems to
indicate that this problem will persist far into the future. The template mechanism makes language interoperability
almost impossible. In general, one must instantiate all templates to be exposed to another language, and remove
templates from public APIs visible in other languages.

The template mechanism, when used to do simple type naming and constant injection, can be quite effective. However, when
used for higher level logical operations and to execute more complicated code rearrangement, there are numerous
problems. The syntax becomes very cumbersome, as in the case of optional template arguments. The logic of instantiation
(type resolution) is opaque, and following the process during debugging is nearly impossible. The gains in source code
size and readability are lost when using templates for more sophisticated code transformation.

\lstset{language=C++,basicstyle=\small,stringstyle=\ttfamily,morekeywords={DistMatrix}}
\paragraph{Examples} The Elemental library~\cite{Elemental2012,elemental-web-page} exhibits two very common uses of
templates for scientific computing. It templates over basic types, but it also uses template markers to switch between
entirely different routines. They are both present in the basic distributed matrix class, \lstinline{DistMatrix},
with declaration:
\begin{lstlisting}
  enum Distribution {
    MC,  // Col of a matrix distribution
    MD,  // Diagonal of a matrix distribution
    MR,  // Row of a matrix distribution
    VC,  // Col-major vector distribution
    VR,  // Row-major vector distribution
    STAR // Do not distribute
  };
  template<typename T, Distribution ColDist,
           Distribution RowDist, typename Int>
  class DistMatrix;
\end{lstlisting}
The first template argument defines the number field over which the matrix operates. This allows identical source to be
used for single precision, double precision, quad precision, and complex matrices, since these types all respond to the
arithmetic operations. At a slightly higher level, search and sort algorithms in the Standard Template Library rely on
the same interface compatibility to write generic algorithms. This can be extended to very high level algorithms, such
as the Conjugate Gradient solver~\cite{saad2003} for sparse linear systems in the \href{http://www.dealii.org/}{dealII}
package~\cite{BangerthHartmannKanschat2007,dealii-web-page}.
\begin{lstlisting}
template <class VECTOR>
template <class MATRIX, class PRECONDITIONER>
void
SolverCG<VECTOR>::solve (const MATRIX         &A,
                         VECTOR               &x,
                         const VECTOR         &b,
                         const PRECONDITIONER &precondition)
{
  if (!x.all_zero()) {
    A.vmult(g,x);
    g.add(-1.,b);
  } else {
    g.equ(-1.,b);
  }
  res = g.l2_norm();

  conv = this->control().check(0,res);
  if (conv) {return;}
  precondition.vmult(h,g);
  d.equ(-1.,h);
  gh = g*h;
  while (conv == SolverControl::iterate) {
    it++;
    A.vmult(Ad,d);
    alpha = d*Ad;
    alpha = gh/alpha;
    g.add(alpha,Ad);
    x.add(alpha,d );
    res = g.l2_norm();
    conv = this->control().check(it,res);
    if (conv != SolverControl::iterate) break;
    precondition.vmult(h,g);
    beta = gh;
    gh   = g*h;
    beta = gh/beta;
    d.sadd(beta,-1.,h);
  }
}
\end{lstlisting}
This code is shared among all implementations of \lstinline!VECTOR!, \lstinline!MATRIX!, and \lstinline!PRECONDITIONER!,
in much the same way it is in OO codes using an abstract base class, similar to PETSc.

However, in complicated numeric codes, it is often the case that template instantiation is substituted for dispatch. For
example, the \lstinline!AlignWith()! method has different implementations depending on the type of the template
arguments. This evaluation of method displatch at compile time avoids the overhead of lookup in a virtual table of
function pointers, but it sacrifces flexibility. With types fixed at compile time, we cannot change types in response to
different input, or new hardware, or simulation conditions without recoding and rebuilding the executable. This makes
exploration of different implementations problematic, particularly in the context of solvers. Moreover, more complex
block solvers for multiphysics systems~\cite{MayMoresi2008} are built out of basic solvers, and runtime type changes
allow construction of a range of powerful solvers~\cite{smcabbkz2012}.

\section{Multi-language Codes}
  \subsection{Python and Wrapping}

\paragraph{Advantages} Multilanguage code allows the designer to combine the strengths of different approaches to
programming. A popular combination in scientific computing is the speed and memory efficiency of languages
like C and Fortran with the flexibility and parsimony of scripting languages such as Python. Python allows inspection of
the full state of the running program, introspection, and management of both memory and variable typing, speeding
development of new code and easing unit testing~\cite{Langtangen09,Nilsen2010}. Python also supports generic programming
since all variables are dynamically typed and do not need to be declared when code is written.

Specialized Python tools have been developed for wrapping C libraries, such as ctypes, SWIG, and Cython. Cython in
particular allows C data structures to be manipulated transparently in Python without copies, Python routines to be
called from function pointers in C, and data conversions to be completely automated. The object structure of C++ can
even be mapped to the object structure in Python. Error and exception handling is also automated. Cython also allows
Python routines to be annotated, and then automatically converted to C and compiled. The \href{http://numpy.scipy.org/}{numpy}
library~\cite{NumpyGuide} allows direct manipulation in Python of multi-dimensional arrays, perhaps using memory
allocated in another language. Operations are performed in compiled code, sometimes on the fly, and without copies,
making it as efficient as standard C, and it can leverage system-tuned linear algebra libraries.

Python string processing and easy data structure manipulation are very useful for managing user input and output. Many
libraries, such as \href{http://www.geodynamics.org/cig/software/pylith}{PyLith}~\cite{PyLithManual,pylith-web-page}, use Python as a top level control
language and then dispatch to C/C++/Fortran for the main numerical processing underneath. Using the tools mentioned
above (PyLith uses SWIG), this process can be almost entirely automated. Moreover, Python's ability to easily expose a
library API, and the use of numpy arrays for data interchange, make it an excellent tool for combining libraries at a
higher level. Libraries solving different physical problems or different models of a given problem can be combined to
attack more complex multi-model, multi-physics, multi-scale problems~\cite{KeyesMcInnesWoodwardEtAl2012,bkmms2012}. In
addition, this wrapping capability has been used to great effect on GPU hardware, incorporating CUDA and OpenCL
libraries into both desktop and parallel computations~\cite{Klockner2012,PyCUDA,PyOpenCL}.

\paragraph{Disdvantages} The central disadvantage for multi-language codes comes in debugging. There are certainly
hurdles introduced into the build system, since different compilation and links steps are needed and many more tests are
needed to verify interoperability, but these can alrgely be handled by standard systems. No tool exists today that can
inspect the state of a running program in the style above, for example Python using a C library. Even the stack trace
after an error is routinely unavailable, although it can be logged by the C library and passed up as is done in
\href{http://code.google.com/p/petsc4py/}{petsc4py}~\cite{petsc4py-web-page}. However, stepping across language boundaries in a debugger is not
yet possible, and this limitation makes debugging new code extremely difficult. Thus, the strategy above works best when
combining several mature single-language libraries, so that debugging is focused only on the interactions between
libraries, which can be seen in the state of the numpy objects communicated among them, rather than on library
internals. Recent developments, including the extension support for Python in gdb 7, indicate that this situation will
improve markedly in the new future.

\lstset{language=python,basicstyle=\ttfamily}
\paragraph{Example} The \href{http://numerics.kaust.edu.sa/pyclaw/}{PyClaw} package~\cite{pyclaw-web-page,alghamdi2011petclaw} combines the
\href{http://www.clawpack.org}{CLAWPACK}~\cite{clawpack45} library for solving hyperbolic systems of partial differential equations on mapped
Cartesian grids with the PETSc~\cite{petsc-user-ref} library parallel linear algebra and scalable solution nonlinear
equations. Incorporation of the PETSc library allowed parallelization of the solvers in both Clawpack and
SharpClaw~\cite{sharpclaw} in only 300 lines of Python, as detailed in~\cite{pyclaw2012}. PETSc parallel data 
structures, in particular the {\bf DA} object for structured grid parallelism, were exposed to Clawpack using Python
numpy arrays as intermediary structures. This allowed no-copy access by both C and Fortran, as well as easy inspection
in the Python code itself. In fact, since numpy structures are used for both wrappers, any PyClaw script can be run in
parallel using the PETSc extension PetClaw simply by replacing the call to \lstinline!import pyclaw! with
\lstinline!import petclaw as pyclaw!. The hybrid code showed excellent weak scaling, when modeling the interaction of a
shock with a low-density bubble in a fluid, on all 65,536 cores of the Shaheen supercomputer at KAUST.

\paragraph{Acknowledgements} The author would like to acknowledge the helpful comments of Wolfgang Bangerth.

\bibliographystyle{plain}
\bibliography{petsc,petscapp}

\end{document}